\documentclass[aps,twocolumn,superscriptaddress,preprintnumbers]{revtex4}
\usepackage{amsmath,amssymb,mathrsfs}
\usepackage{subfigure}
\usepackage[dvips]{graphicx}
\usepackage{hyperref}
\usepackage{paralist}
\usepackage{epstopdf}
\usepackage{dcolumn}
\usepackage{color}

\def\be{\begin{equation}}
\def\ee{\end{equation}}
\def\bea{\begin{eqnarray}}
\def\eea{\end{eqnarray}}

\def\tbf{\textbf}

\def\up{{\uparrow}}
\def\down{{\downarrow}}

\begin{document}
\title{Topological orbital ladders}

 \author{Xiaopeng Li} \affiliation{Department of Physics
  and Astronomy, University of Pittsburgh, Pittsburgh, Pennsylvania}

\author{Erhai Zhao}
\affiliation{School of Physics, Astronomy, and Computational Sciences, George Mason University, Fairfax, Virginia}

\author{W. Vincent Liu\footnote{email: w.vincent.liu@gmail.com}}
\affiliation{Department of Physics and Astronomy, University of
Pittsburgh, Pittsburgh, Pennsylvania}
\affiliation{Center for Cold Atom Physics, Chinese Academy of Sciences,Wuhan 430071, China}

\date{\today}


\maketitle {\bf Synthetic quantum orbital materials, such as the
  recent double-well optical lattices loaded with 
  $s$ and $p$ orbital
  atoms~\cite{2011_Wirth_pband,2011_Sengstock_spin-dependent-dw_NPHYS},
  open an avenue to exploit symmetries beyond natural
  crystals.  Exotic superfluid states were reported for 
bosons.
  Here, we unveil a topological phase of interacting fermions on
  a two-leg ladder of unequal parity orbitals, derived from the
  experimentally realized double-well lattices by dimension
  reduction. $Z_2$ topological invariant originates simply from the
  staggered phases of $sp$-orbital quantum tunneling, requiring none
  of the previously known mechanisms such as spin-orbit coupling or
  artificial gauge field.  Another unique feature is that upon
  crossing over to two dimensions with coupled ladders, the edge
  modes from each ladder form a parity-protected flat band at zero
  energy, opening the route to strongly correlated states controlled
  by interactions.  Experimental signatures are found in density
  correlations and phase transitions to trivial band and Mott
  insulators.}

\begin{figure}[htp]
\begin{center}
\includegraphics[angle=0, width=\linewidth]{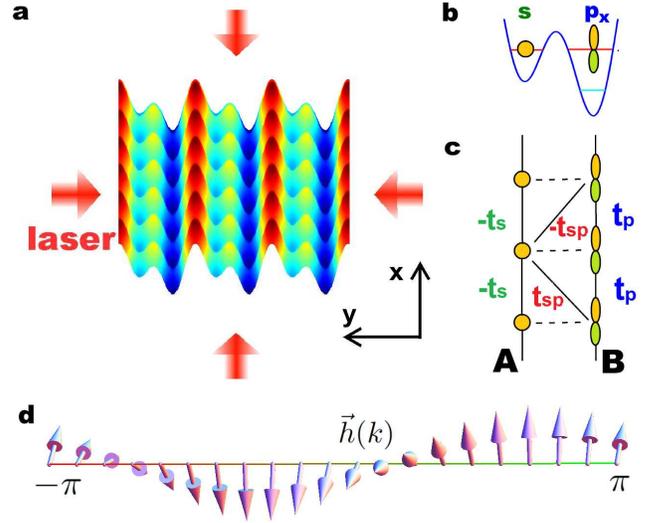}
\end{center}
\caption{The uneven  $sp$-orbital ladder system made from a
    two-dimensional double-well optical lattice through dimension
    reduction. 
\textbf{a}, An optical lattice of uneven sub-wells (light and dark blue), with parameters
$V_x/V_{1} = 0.3$, $V_{2}/V_{1} = 1$ and $\phi = 0.6 \pi$, develops
high barriers (red ridges) in the $y$ direction, slicing  the lattice 
into dynamically decoupled uneven two-leg ladders.
\textbf{b} and \textbf{c}, Schematic side and top views,
respectively, of the ladder illustrate tunneling ($t$'s) of fermions prepared
in the degenerate $s$ and $p_x$ levels.
{\textbf{d}, Topological winding of Hamiltonian across the Brillouin zone.} } 
\label{fig:exp-setup}	
\end{figure}

The uneven double-well potential formed by laser light has been developed into a powerful tool for 
quantum gases by numerous 
groups~\cite{2006_NIST_Lundblad_double-well_pra,2007_NIST_double-well_Nature,2008_Bloch_Trotzky_Double-well_Science,2011_Wirth_pband,2011_Sengstock_spin-dependent-dw_NPHYS,2011_Sengstock_honeycomb_BEC}. 
For instance, 
{ controls of atoms on the $s$- and $p$-orbitals of the
checkerboard~\cite{2011_Wirth_pband} and 
hexagonal~\cite{2011_Sengstock_honeycomb_BEC} optical lattices
have been demonstrated, and correlation between these orbitals 
tends to give exotic quantum states~\cite{2011_Sengstock_honeycomb_BEC,2011_Zhou_doublewell_PRB}.} Motivated by these developments, 
we consider a lattice of uneven double-wells 
where fermionic atoms are loaded up to the $s$- and $p$-orbital levels
of the shallow and deep wells respectively, as shown in
Fig.~\ref{fig:exp-setup}.  The spatial symmetry of the orbital wave
function dictates the complex hopping amplitudes between nearby sites.
Under certain circumstances, as for the uneven double wells, the
orbital hopping pattern is sufficient for producing topologically
nontrivial band structures~\cite{2011_Sun_TSM}.

We will first focus on a one-dimensional
ladder system illustrated in Fig.~\ref{fig:exp-setup}\textbf{b} and \ref{fig:exp-setup}{\bf c}. 
This corresponds to the quasi-one dimensional limit of a 
standard double-well optical lattice,
with the optical potential given by
\[
V(x,y) = V_x \sin^2 (k x)  + V_{1} \sin^2 (k y) 
  + V_{2} \sin^2 (2 k y+\frac{\phi}{2} ).
\]
This optical lattice has a double well structure in the $y$-direction.  
For $V_{1,2}\gg V_x$, there is a large tunneling
barrier between double wells in the $y$-direction, so in low
  energy physics the 
two dimensional system decouples  into an array of
dynamically isolated two-leg ladders of  $A$ and $B$ sub-wells 
(Fig.~\ref{fig:exp-setup}), with each ladder extending  in the $x$-direction. 
The relative well depth of the two legs is controlled by the phase 
$\phi$ and further by the ratio $V_{2}/V_{1}$. 
We will focus on a situation, similar to the 
setup in the experiment~\cite{2011_Wirth_pband}, where the $s$-orbital 
of leg $A$ has roughly the same energy as the $p_x$-orbital of leg $B$ (other $p$-orbitals have
much higher energy). 
{For example, one can choose $V_1 = 40 E_R$, $V_2 = 20 E_R$, $V_x = 3.8E_R$ and $\phi = 0.9 \pi$ in 
experiments, where $E_R$ is the recoil energy ${\hbar^2
  k^2}/{2m}$, with $m$ the mass of atom and $k$ the wave number of the laser. 
Such a setup will give the $A$ ($B$) wells a depth $2.7E_R$ ($8.1E_R$).
The lattice constant $a = {\pi}/{k}$ will be set as the length unit in this paper. 
We now consider a single species  of  fermions 
occupying these orbitals, with the low-lying $s$-orbital of leg $B$ 
completely filled 
(alternatively fermions can be directly 
loaded into the $p_x$-orbital of leg $B$ by techniques developed for bosons  
in Ref.~\cite{2007_NIST_double-well_Nature,2011_Wirth_pband}).

The Hamiltonian of the $sp$-orbital ladder is then given by
\bea
\textstyle H_0= \textstyle \sum_j
                C_j ^\dag
                \left[
                \begin{array}{cc}
                -t_s & -t_{sp} \\
                t_{sp} & t_p
                \end{array}
                \right]
                C_{j+1}
                +h.c. 
       - \sum_j \mu    C_j ^\dag C_j,
\label{eq:ham}
\eea
where $C^\dagger_j = \left[ a_s^\dagger (j) , a^\dagger_{p_x} (j) \right]$, 
$a_s^\dagger (j) $ and $a^\dagger_{p_x} (j) $ are the fermion creation operators 
for the $s$- and $p_x$-orbitals on the A and B leg respectively. The
relative sign of the hopping amplitudes is fixed by parity symmetry of the
$s$ and $p_x$ orbital wave functions. As depicted in Fig.~\ref{fig:exp-setup}\textbf{c},
the hopping pattern plays a central role in producing a $Z_2$ topological phase. 
With a proper global gauge choice, $t_s$, $t_p$ and $t_{sp}$ are all positive.
The rung index $j$ runs from $0$ to $L-1$ with $L$ the system size.
We consider half filling (one particle per unit cell), for which the chemical 
potential $\mu =0$, and the Hamiltonian is particle-hole symmetric under transformation 
$C_j \to (-1)^j C_j ^\dag$.
This $sp$-orbital ladder contains a staggered hopping, which
  is analogous to 
  the Su-Schrieffer-Heeger model for the dimerized 
  polyacetylene, an electron-conducting polymer~\cite{1979_Su_SSH-model_PRL}. 
Its physics is also connected to the more familiar frustrated ladder with 
magnetic flux~\cite{2012_Jaefari_Fradkin_PDWSC_PRB}, but the
$sp$-orbital ladder {appears much
easier to 
realize experimentally.}

In the momentum space the Hamiltonian takes a simple and suggestive form,
\be
\textstyle \mathcal{H} (k ) = h_0 (k) \mathbb{I} + \vec{h}(k) \cdot \vec{\sigma},
\label{eq:bulkhk}
\ee
where $h_0 (k) =(t_p -t_s) \cos(k)$,
$h_x =0$,
$h_y (k) =  2t_{sp} \sin(k)$ and
$h_z (k)  = -(t_p + t_s) \cos(k)$. Here, $\mathbb{I}$ is the unit matrix,  
$\sigma_x$, $\sigma_y$ and $\sigma_z$ are Pauli matrices in the two-dimensional
orbital space. The energy spectrum consists of two branches,
\be
\textstyle E_{\pm} (k)  = h_0 (k)
    \pm \sqrt{h_y^2 (k) + h_z ^2 (k) },
\label{eq:bulkspectra}
\ee
with a band gap $E_g  = \min(2t_p+2t_s, 4t_{sp} )$, which
closes at either $t_{sp} =0$ or $t_s +t_p =0$. An interesting limit that
highlights the nontrivial band structure of our model is that
when $t_p = t_s = t_{sp}$, the two bands are both completely flat.
To visualize the topological properties of the 
band structure, one notices that as $k$ is varied
from $-\pi$ through 0  to $+\pi$, crossing the entire Brillouin zone, 
the direction of the $\vec{h}(k)$ vector winds an angle of
  $2\pi$ (Fig.~\ref{fig:exp-setup}\textbf{d}).
The corresponding Berry phase is half of the angle, $\gamma=\pi$. 
For a one-dimensional Hamiltonian with particle-hole symmetry, the Berry phase
is quantized~\cite{2008_Qi_TFT_TI}, 
$\gamma \mod 2\pi=0$ or $\pi$, which defines the trivial
and Z$_2$ topological insulator, respectively.
The $sp$-orbital ladder exhibits $\gamma =\pi$ for $t_{sp} \neq 0$, so it is
a $Z_2$ topological insulator
at half filling. This is verified by explicit calculation of the Berry
phase (see Supplementary Information).

The nontrivial topology of the ladder system also manifests in existence of edge 
states. It is easiest to show the edge states in the flat band limit,
$t_s = t_p = t_{sp}\equiv t$, by introducing {auxiliary operators, }
$\phi_{\pm}(j)=[a_{p_x}(j)\pm a_s(j)]/\sqrt{2}$. Then the Hamiltonian
only contains coupling between $\phi_+$ and $\phi_-$ of nearest neighbors,
but not among the $\phi_+$ (or $\phi_-$) modes themselves,
\be
\textstyle H_0 \rightarrow 2t\sum_j \phi_-^\dagger(j)\phi_+(j+1) + h.c.
\ee
Immediately, one sees that the operators $\phi_+ (0)$
and $\phi_-(L-1)$ at the left and right end are {each} dynamically isolated from the bulk,
and do not couple to the rest of the system 
(Fig.~\ref{fig:eigenmodes}\textbf{a}). These loners describe the two
edge states at zero energy.

{ It is instructive to map} the $sp$-orbital ladder to 
two decoupled Majorana chains~\cite{2001_Kitaev_Majorana_Phys-usp}.
For each rung $j$, we introduce 4 Majorana fermion operators,
$\psi_{1 \pm}= (\phi_\pm + h.c.)$ and 
$\psi_{2 \pm}= (i\phi_{\mp}+ h.c.)$. Then the Hamiltonian
is transformed into
\be
\textstyle H_0 \rightarrow it \sum_j \left[
	  \psi_{2 +}  (j) \psi_{1 +}(j+1)
	  -\psi_{1-} (j)  \psi_{2 -} (j+1) \right],
\label{eq:ham-majorana}
\ee
for which four {unpaired Majorana fermions} $\psi_{1+} (0)$, $\psi_{2-}(0)$, $\psi_{2+} (L-1)$ 
and $\psi_{1-}(L-1)$ emerge.
The four unpaired Majorana fermions manifest a Z$_2$ (Z$_4$) degenerate ground state 
in the canonical (grand-canonical) ensemble at half filling.  
Note that the Majorana number as defined in Ref.~\cite{2001_Kitaev_Majorana_Phys-usp} 
is $1$ for 
our double Majorana chains. While on general grounds coupling between
the four unpaired Majorana fermions is allowed, local coupling, e.g., in the form of 
$i\epsilon \psi_{1+} (0) \psi_{2-} (0) = -2\epsilon \left(C_0 ^\dag\left[\sigma_0 + \sigma_x\right] C_0\right) + const$, 
is however prohibited in our system due to the required particle-hole symmetry.

Once the particle-hole symmetry is broken, topologically protected Majorana fermions with Majorana number $-1$~\cite{2001_Kitaev_Majorana_Phys-usp} can be realized on the $sp$-ladder using schemes similar to those proposed in 
Ref.~\cite{2011_Jiang_Majorana_PRL,2010_Lutchyn_Majorana_PRL,2010_Oreg_Majorana_PRL}, e.g., 
by inducing weak pairing of the form   
$\sum_j \Delta a_s (j) a_p (j)+h.c.$.  In the parameter
regime $2t_s <|\mu| <2 t_p$, the Majorana number is $-1$ and the resulting Majorana zero 
modes are topologically protected. 
{The topologically protected Majorana state is a promising candidate 
for topological quantum computing~\cite{2010_Sau_Majorana_PRL,2011_Alicea_Fisher_QC}.} 
Compared to previous proposals ~\cite{2011_Jiang_Majorana_PRL,2010_Lutchyn_Majorana_PRL,2010_Oreg_Majorana_PRL},
the present ladder system does not require spin-orbit coupling.

Away from the flat band limit, in general the wave functions of the edge states
are no longer confined strictly at $j=0$ or $L-1$, but instead decay
exponentially into the bulk with a characteristic length scale
$\xi = 2/\log(\left| (\sqrt{t_s t_p} + t_{sp})/(\sqrt{t_s t_p} - t_{sp}) \right|)$.
The analytical expression for the edge
state wave functions in the general case are discussed in Supplementary Information.
At the critical point $t_{sp} =0$, the bulk gap closes and $\xi \to \infty$.
The existence of zero energy edge states is also confirmed by numerical calculation 
as shown in Fig. \ref{fig:eigenmodes}\textbf{b} and \ref{fig:eigenmodes}\textbf{c}.

\begin{figure}[htp]
\begin{center}
\includegraphics[angle=0,width=1.0\linewidth]{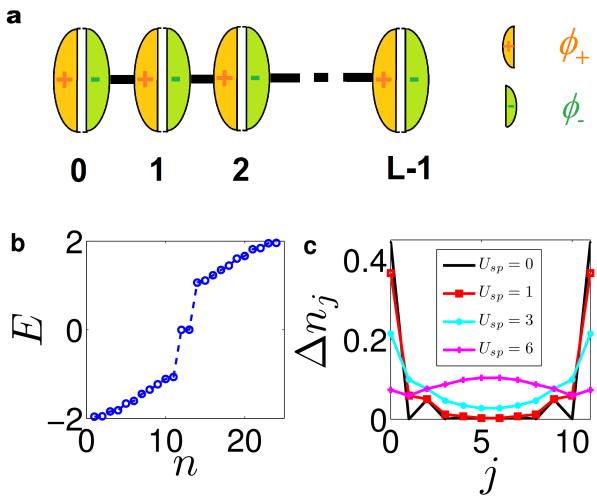}
\end{center}
\caption{
    Bulk and edge eigenstates of the $sp$-orbital ladder.
    \textbf{a}. A pictorial representation of the simplified Hamiltonian in the
    flat band limit $t_s =t_p = t_{sp}$ showing the emergence of isolated edge modes. 
    The definition of the $\phi_{\pm}$ operator
    is given in the main text. \textbf{b}. The eigen energy
    of a ladder with finite length $L=12$ showing two degenerate zero
    energy states inside the gap. 
    \textbf{c}. The probability distribution of the in-gap states (Eq.~(\ref{eq:n_j})) for
    varying strengths of inter-orbital interaction $U_{sp}$. The in-gap states are
    {shown localized} on the edges and survive against finite
    interaction. In \textbf{b} and \textbf{c}, we choose $t_s = t_p = 2 t_{sp}$ (taken as the energy unit). 
}
\label{fig:eigenmodes}
\end{figure}

For a finite ladder of length $L$ with open boundary condition and populated by
$L$ fermions (half filling), $L-1$ fermions will occupy the valence
band (bulk states) and one fermion will occupy the edge states
 (Fig.~\ref{fig:eigenmodes}). Since 
the two edge states are degenerate, the ground state has a double (Z$_2$)
degeneracy. The edge state is a fractionalized object
carrying half charge (cold atoms are charge neutral, here charge refers to
the number of atoms).
This becomes apparent if we break the particle-hole symmetry by going infinitesimally away 
from half filling, e.g., tuning chemical potential
$\mu = 0^+$. Then, the valence bands and the two edge states will be 
occupied. 
With a charge density distribution on top of the half 
filled background defined as $\rho (j) = C ^\dag_jC_j  - 1$,  one finds
$\sum_{0\le j<d} \langle \rho (j) \rangle|_{\mu = 0^+}
  =\sum_{L-d-1< j'<L} \langle \rho (j') \rangle|_{\mu = 0^+}
  = \frac{1}{2}$, where $d$ satisfies $\xi\ll d\ll L$ { (e.g., take $d=5 \xi$)}.
A characteristic feature of the $Z_2$ topological insulator (with the number of atoms fixed) 
is the topological
anti-correlation of the charge at the boundaries,
\[
\lim_{L\to \infty} \sum_{ j, j'} 
    \langle \rho(j) \rho (j') \rangle = - \frac{1}{4}. 
\]  
In the flat band limit, $\xi \to 0$, the edge states are well 
localized at the two ends of the ladder. The topological anti-correlation simplifies as 
$\langle \rho(0) \rho(L-1) \rangle = -\frac{1}{4}$, and the half charge 
is also well localized, i.e.,
$\langle \rho (0) \rangle _{\mu =0^+} 
  = \langle \rho (L-1) \rangle _{\mu =0^+}  = \frac{1}{2}$.
Since the edge states are well isolated from the bulk
states by an energy gap, they are stable against local Gaussian fluctuations. 
The coupling between the two edge states
vanishes in the thermodynamic
limit ($L\to\infty$), because the hybridization induced gap scales as $\exp(-\alpha L)$ as
$L \to \infty$~\cite{2009_Kitaev_TopoPhase}. 

An interesting topological phase transition to a trivial insulator 
can be tuned to occur when rotating the atoms on the individual sites, 
for example, by applying the technique demonstrated in Ref.~\cite{2010_Gemelke_lattice-Rotating_arXiv}. 
Such an 
individual site rotation amounts to addition of an imaginary transverse (along $y$) 
tunneling between $s$- and 
$p_x$-orbitals in our Hamiltonian, $\delta H = \Delta_y \sum_j C_j ^\dag  \sigma_y C_j$.
This term preserves particle-hole symmetry
but breaks both parity and time-reversal symmetries. 
The total Hamiltonian in the momentum space now reads
$ \mathcal{H}' (k) = \mathcal{H} (k) + \Delta_y \sigma_y$.
The quantized Berry phase remains at $\pi$ for
$\Delta_y$ {smaller than} the critical value $\Delta_y ^c = 2 t_{sp}$,
so the topological insulator phase is stable
against finite $\Delta_y$. This further shows that the Z$_2$
topological phase is protected by particle-hole symmetry.
For $\Delta_y$ {greater than} $\Delta_y^c$, 
Berry phase vanishes and the system becomes a trivial band 
insulator. At the critical point the band gap
closes. 
Apart from the
Berry phase, the topological distinction between $\mathcal{H}' (k) $ and $\mathcal{H} (k)$
can also be seen from a gapped interpolation~\cite{2008_Qi_TFT_TI} as
discussed in Supplementary Information.
Besides probing the half charges on the boundaries, another signature
  for 
the critical point of the topological phase
transition is the local density fluctuation,
$\delta \rho = \frac{1}{L} \sum_j \sqrt{\langle \rho(j) \rho(j)\rangle }$.
$\delta \rho$ is $1/\sqrt{2}$ when $\Delta_y =0$, {independent of other} parameters $t_s$, $t_p$ and $t_{sp}$,
and decreases monotonically with increasing $\Delta_y$. The peaks of
${d \delta \rho^2}/{d \Delta_y}$ {reveal the critical
points}~(Fig.~\ref{fig:ti-trivial-transition}\textbf{b}) and provide
a reliable tool of detecting the topological phase transition in experiments.

\begin{figure}[htp]
\begin{flushleft}
\includegraphics[angle=0,width=1.\linewidth]{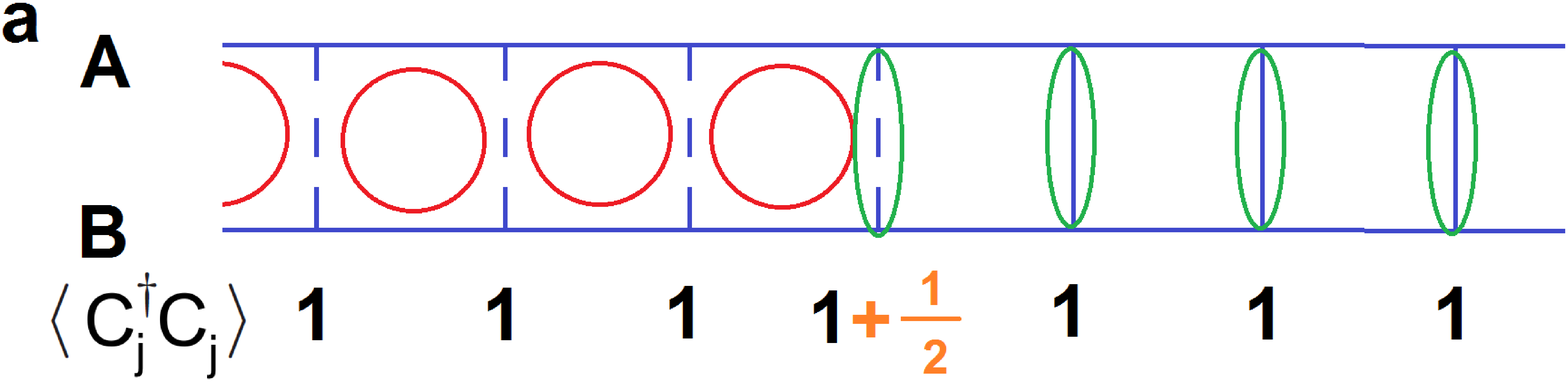}
\includegraphics[angle=0,width=0.9\linewidth]{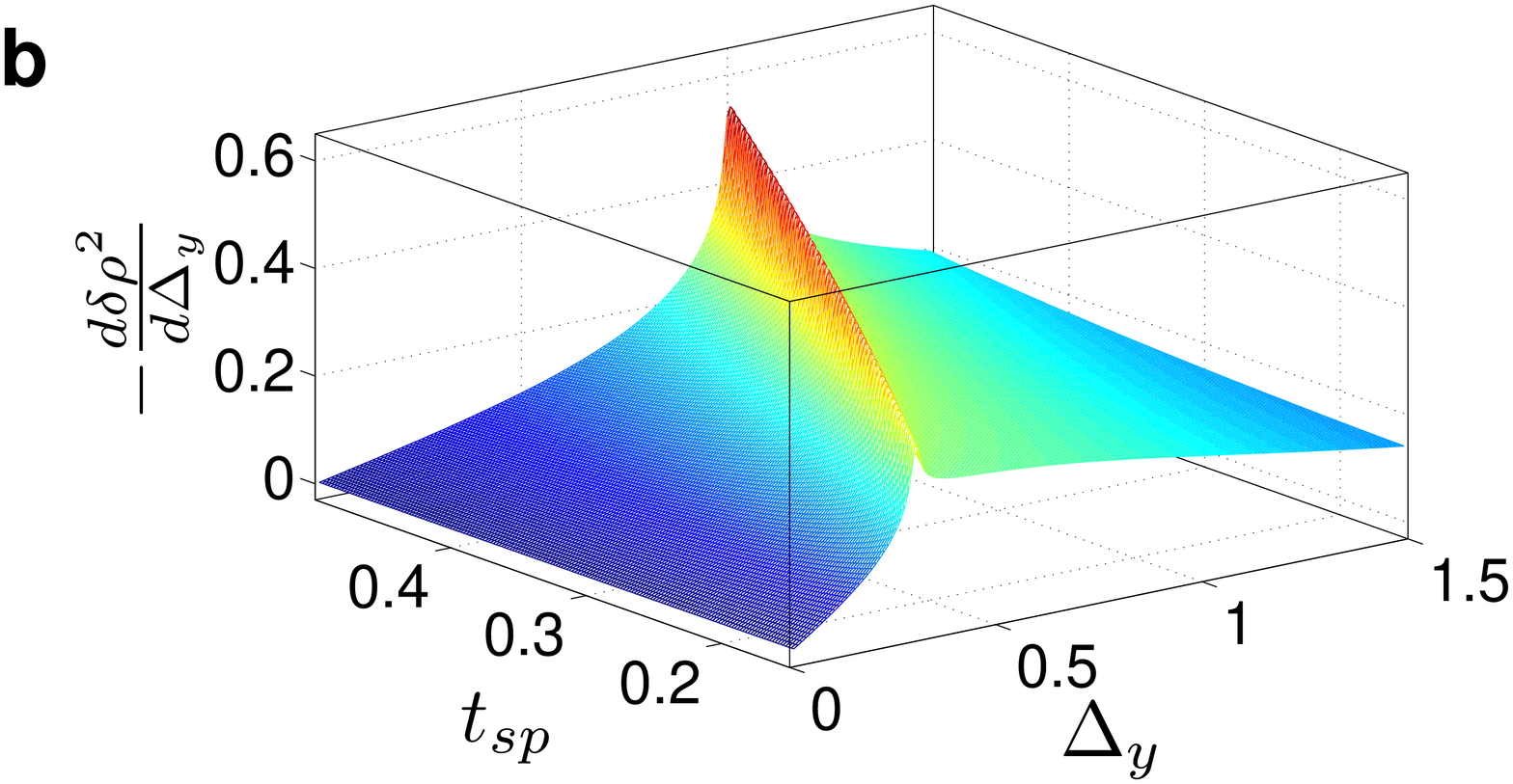}
\end{flushleft}
\caption{Phase transition between the topological and the trivial
    band insulator.  \textbf{a}. A domain wall between
    a topological insulator ($t_s = t_p = t_{sp}$, $\Delta_y =0$, left)
    and a trivial insulator ($t_s = t_p = t_{sp} =0$, right).
    The circle represents the delocalized fermion shared by two neighboring
    rungs as depicted in Fig. \ref{fig:eigenmodes}{\bf a},
whereas the ellipse represents localized fermion without hopping.
The additional charge $\frac{1}{2}$ in the middle is the fractional
charge carried on the domain wall. 
\textbf{b}. The derivative of density fluctuation, $-\frac{d \delta \rho^2}{d \Delta_y}$. 
It develops sharp peaks, measurable in experiments, along the line of
topological critical points. }
\label{fig:ti-trivial-transition}
\end{figure}

It is feasible to prepare the ladder with phase separation: e.g., 
a topological insulator on the left half but a trivial insulator on the 
right half. This can be achieved by 
rotating the lattice sites on half of the ladder only.
The system is now described by 
\be
\textstyle H_\eta = H
+ \frac{ {\Delta}_y}{2} \textstyle \sum_j \left[1-\cos  \eta(j)  \right]
    C_j^\dag \sigma_yC_j
\label{eq:ham-theta}
\ee
with a field configuration $\eta(j)$ which satisfies the boundary conditions
$\eta (j =-\infty) = 0$ and $\eta (j = + \infty ) = \pi$, and
${\Delta}_y > \Delta_y ^c$. The charge
distribution induced by the domain wall (the phase boundary) is calculated both numerically and
from effective field theory shown in the Supplementary Information. Both
approaches {cross-verify that} the domain wall carries half
charge~(Fig.~\ref{fig:ti-trivial-transition}\textbf{a}). The half charge can be detected
by the single site imaging technique 
in experiments~\cite{2010_Bakr_singlsite-image_Science,2010_Sherson_singlsite-image_Nature}.

\begin{figure}[htp]
\begin{center}
\includegraphics[angle=0,width=0.95\linewidth]{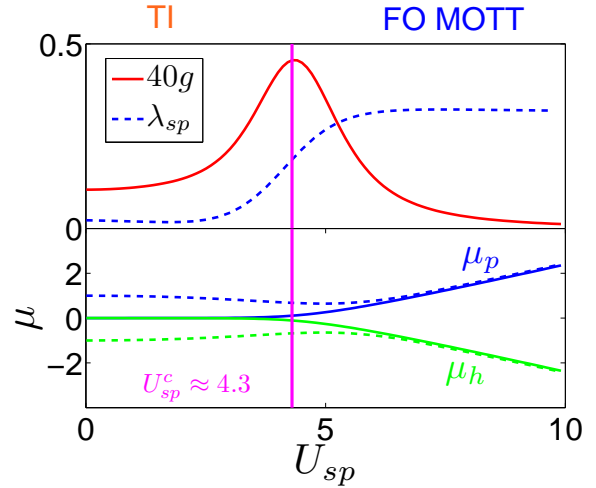}
\end{center}
\caption{ Transition from a topological insulator (TI) phase into
a Mott insulator with ferro-orbital (FO) order with increasing interaction. 
Top panel shows the fidelity metric $g$ and 
the ferro-orbital order parameter $\lambda_{sp}$. 
 Bottom panel shows the particle/hole chemical 
potential ($\mu_p$/$\mu_h$). 
{ The finite charge gap 
$\mu_p - \mu_h$ in the bulk calculated with periodic boundary condition (dashed lines)
comparing with the vanishing gap with open boundary condition (solid lines) indicates in-gap states 
on the edge. The length $L$ is $12$, and 
$t_s = t_p =2 t_{sp}$ (taken as the energy unit) in this plot. }
 }
\label{fig:fermion_phasetransition}
\end{figure}

We further examine the stability of the topological phase
and its quantum phase transitions in the presence of interaction
using exact diagonalization. For { single-species} 
fermions on the $sp$-orbital ladder, the leading interaction term is
the on-site repulsion between different orbitals,
\be
H_\text{int}= \sum_j U_{sp} \left[ n_s (j) - \frac{1}{2} \right] 
    \left[n_p (j) -\frac{1}{2} \right]. \label{eq:Vsp}
\ee
We compute the fidelity metric $g$ as function of the interaction strength $U_{sp}$ 
(see the Methods section). 
A peak in the fidelity metric indicates a quantum phase 
transition~\cite{2011_Varney-Sun_ITI_PRB}. Our numerical results
(Fig.~\ref{fig:fermion_phasetransition}) show that  
the topological phase is stable for $U_{sp} <U_{sp} ^c$, with
robust in-gap (zero-energy) states~\cite{2011_Varney-Sun_ITI_PRB} 
localized on the edges
(Fig.~\ref{fig:eigenmodes}\textbf{c}). 
For stronger interaction the ladder undergoes a quantum phase transition to 
a Mott insulator phase, which exhibits ferro-orbital order with
order parameter defined as $\sqrt{\lambda_{sp}}=\langle C_j ^\dag \sigma_x  C_j \rangle$.

\begin{figure}[htp]
\begin{center}
\includegraphics[angle=0,width=1.1\linewidth]{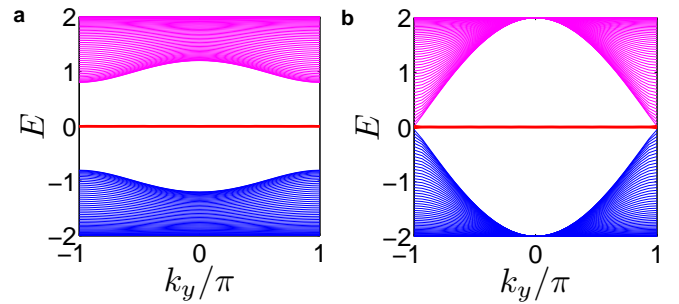}
\end{center}
\caption{ The energy spectra of the two dimensional system of coupled 
ladders with $t_s = t_p = 2t_{sp}$ { (taken as the energy unit here), and 
length $L = 200$.} 
An open (periodic) boundary condition 
is applied in the $x$ ($y$) direction. \textbf{a} and \textbf{b} show the spectra with the small 
and large inter-ladder coupling, $t_{sp}'=t_{sp}/5$ and  $t_{sp}' = t_{sp}$, respectively.
A flat band (red line) at zero energy with double degeneracy generically appear for $0<t_{sp}'<t_{sp}$.
}
\label{fig:2d_flat_edge}
\end{figure}

Finally we point out a remarkable difference from the
  Su-Schrieffer-Heeger model. The zero-energy states 
in the present $sp$-ladder survive even when the system 
is extended to two dimensions with finite inter-ladder coupling (e.g., by reducing
$V_{1,2}$ relative to $V_x$ in the
setup of Fig.~\ref{fig:exp-setup}{\bf a}). The zero modes of individual ladder 
morph into a flat band with double degeneracy (Fig.~\ref{fig:2d_flat_edge}). The lack
of dispersion in the $y$ direction is related to the inter-ladder hopping pattern, 
which does not directly couple the edge states but only $s$- and $p$-orbitals
on different rungs. The unexpected flat band in 2D is an exact consequence of  the $p$-orbital parity 
and hence is protected by symmetry. The flat dispersion can be rigorously proved using an
unitary transformation and arguments based on the quantization of 
Berry phase in the Supplementary Information. 
{The flat band is 
reminiscent of that at the zigzag edge of graphene, but with the difference 
that the present flat band is protected by the parity of the orbital wavefunctions.
The diverging density of states associated with the flat band provides a fertile ground 
for interaction-driven many body instabilities. Future work will tell whether
strongly correlated topological states exist in such two-dimensional 
interacting system.

\section*{Methods}
\paragraph*{\bf Imaginary transverse tunneling.}
By rotating individual lattice sites the induced
bare coupling term is $\sum_j \Omega \hat{L}_z (j)$ where the angular
momentum operator
$\hat{L}_z=-i(a^\dag_{p_y} a_{{p_x}} -a^\dag_{p_x} a_{{p_y}})$. This
term couples the $p_x$ to $p_y$-orbitals of the $B$-leg. 
One can tune the rotating
frequency to match $\Omega$ with the transverse tunneling
$t_{sy}$ from the $p_y$-orbital of the $B$-leg to
the nearby $s$-orbital on the $A$-leg.
Despite the large energy band gap ($\epsilon_y$) which separates the
$p_y$-orbitals from the degenerate  $s$ 
and $p_x$  orbitals (bare in mind the $s$ and $p_x$ orbitals are from
different legs of the ladder), 
the low energy effective Hamiltonian receives a standard 2nd-order
effect from virtual processes,
in which a particle jumps from a $p_x$-orbital to the on-site $p_y$-orbital
and then to the nearby $s$-orbital.
The correction is given by
$\frac{\Omega^2}{\epsilon_y} C_j ^\dag \sigma_y C_j$,
which makes an imaginary transverse tunneling between nearby $s$- and
$p_x$-orbitals.

\paragraph*{\bf Interaction effects.}
To characterize the stability of the topological phase against the 
{ inter-orbital interaction $H_\text{int}$ (see Eq. \eqref{eq:Vsp}),} 
we use the exact diagonalization method to 
calculate the fidelity metric 
$
g  \equiv 2\left[ 
 1- |\langle \psi_L ^0 (U_{sp})| \psi_L ^0 (U_{sp}+ \delta U_{sp}) \rangle |\right]
/ L (\delta U_{sp} )^2
$,
where $|\psi_L ^0 (U_{sp}) \rangle $ is the ground state wave function
of the Hamiltonian $H=H_0+ H_\text{int} $ for a finite chain of length $L$. 
In presence of interaction, the edge states 
survive as in-gap states (zero energy single particle/hole excitations)~\cite{2011_Varney-Sun_ITI_PRB}. 
The energy of single particle (hole) excitation is defined as
$\mu_p$ ($\mu_h$) by $\mu_p = E_{L+ 1}  - E_{L} $ ($\mu_h = E_{L} - E_{L-1}$),
where $E_{N}$ is the ground state energy of the ladder loaded with $N$ fermions.
The spatial distribution of the in-gap states is defined as
the density profile ($\Delta n_j$) of a hole created out of the ground state, 
which is
\be
\Delta n_j
    = \langle \psi^0 _{L} |\rho(j) |\psi^0 _{L} \rangle
    - \langle \psi^0_{L-1}|  \rho(j) | \psi^0_{L-1} \rangle,
\label{eq:n_j}
\ee
where $|\psi_N ^0\rangle$ is the ground state with $N$ fermions.
The density profile $\Delta n_j$ is found to be localized on the edges
when $U_{sp} \ll U_{sp} ^c$, 
and to delocalize when approaching the critical point and finally
disappear (Fig.~\ref{fig:eigenmodes}\textbf{c}).  
The Mott state appearing at the strong coupling regime has a ferro-orbital order  
$\langle \hat{O}_{sp} (j) \rangle$, with 
$\hat{O}_{sp} (j) = C_j ^\dag \sigma_x  C_j$ (Fig.~\ref{fig:fermion_phasetransition}).   
In our numerical { calculation of finite system size,} the correlation matrix 
$[{\cal C}]_{j_1 j_2} =\langle \hat{O}_{sp} (j_1) \hat{O}_{sp} (j_2)\rangle$ is calculated and 
the strength of the ferro-orbital order $\lambda_{sp}$ is defined as the maximum 
eigenvalue of $[{\cal C}]/L$, extrapolated to the thermodynamic limit.

\section*{Acknowledgment}
The authors acknowledge helpful discussions with I. Bloch, 
E. Fradkin, X. J. Liu, S. Das Sarma, and S. Tewari.  
This work is supported by A. W. Mellon Fellowship (X.L.), AFOSR
(FA9550-12-1-0079) (E.Z. and W.V.L.), ARO (W911NF-11-1-0230) and
ARO-DARPA-OLE (W911NF-07-1-0464) (X.L. and W.V.L.), the National Basic
Research Program of China (Grant No 2012CB922101), and Overseas
Scholar Program of NSF of China (11128407) (W.V.L.).

\section*{Author Information}
The authors declare no competing financial interests.
Correspondence and requests for material should be sent to w.vincent.liu@gmail.com.
Supplementary information accompanies this paper.

\begin{widetext}

\renewcommand{\thesection}{S-\arabic{section}}
\renewcommand{\theequation}{S\arabic{equation}}
\setcounter{equation}{0}  
\renewcommand{\thefigure}{S\arabic{figure}}
\setcounter{figure}{0}  

\section*{\Large\bf Supplementary Information}

\section{\bf Topological index}
\label{sec:top-index}
The topological nature of the $sp$-orbital ladder can be understood
in terms of the winding number of the Hamiltonian in the momentum space.
Given a Hamiltonian
\bea
\mathcal{H} (k ) = h_0 (k) \mathbb{I} + \vec{h}(k) \cdot \vec{\sigma},
\label{eq:Shamk}
\eea
with $h_x =0$, the winding number is defined as
\bea
W = \oint \frac{dk}{4\pi}
\epsilon_{\nu \nu'} \hat{h}_{\nu} ^{-1}  (k) \partial_k \hat{h}_{\nu'},
\eea
where $\hat{h} = \frac{\vec{h}}{|\vec{h}|}$ and $\epsilon_{yz} = -\epsilon_{zy}=1$.
This winding number is $1$ for the $sp$-orbital ladder in the topological insulator phase
(Fig.~\ref{fig:exp-setup}\textbf{d}). 

The Hamiltonian in equation~(\ref{eq:Shamk}) can be written as
$\mathcal{H} (k) = U(\mathbb{I} + \sigma_z ) U^\dag$,
where $U = \exp \left(  i\sigma_x \theta (k)/2  \right)$, with
$\theta(k)$ defined by
$
\left[
\begin{array}{c}
\cos(\theta) \\
\sin(\theta)
\end{array}
\right]
=\left[
\begin{array}{c}
 \hat{h}_z (k) \\
 \hat{h}_y (k)
 \end{array}
 \right]
$.
The eigenvector of the lower branch is given as
$u(k) = e^{-i\theta(k)/2 } U
\left[
    \begin{array}{c}
    0 \\
    1
    \end{array}
\right]$.
The Berry phase is given by
\bea
\gamma &=& \oint dk i u^\dag \partial_k u \nonumber \\
    &=& i\oint dk  \left[ e^{-i\sigma_x \theta/2} e^{i\theta/2 }
                        \partial_k \left( e^{i \sigma_x \theta/2} e^{-i\theta/2 } \right)
                        \right]_{22} \nonumber \\
    &=& \frac{1}{2} \oint dk \partial_k \theta
        = \frac{1}{4} \oint dk
            \frac{ \partial_k \sin(\theta)}{\cos(\theta)}
            -\frac{\partial_k \cos(\theta) }{\sin(\theta) } \nonumber \\
     &=& -W \pi.
\eea
With the eigenvector $u(k)$ multiplied by a phase factor
$e^{i\phi(k)}$ in which $\oint dk \partial_k \phi(k) = 2n\pi$,
the Berry phase changes $2n\pi$. This means the Berry phase
$\gamma = W\pi \mod 2\pi$. The winding number $W$ being even and odd defines two classes
of topological states in one dimension. 

\section{\bf The wave functions of edge states}
\label{sec:wave-edgestate}
To see the existence and the property of of the edge states away from the flat band 
limit it is useful to rewrite $H_0$ in the following form
\be
\textstyle H_0 = \sum_j
\Phi_j ^\dag
    \left[
    \begin{array}{cc}
    t_1 & t_2 \\
    t_3 & 0
    \end{array}
    \right]
\Phi_{j+1}
+h.c.,
\label{eq:sup-ham}
\ee
with
\bea
&&
\textstyle \Phi_j  =
    \left[
    \begin{array}{c}
    \phi_+ (j) \\
    \phi_- (j)
    \end{array}
    \right]=
\frac{1}{\sqrt{t_s + t_p}} \left[
\begin{array} {cc}
    \sqrt{t_s} & \sqrt{t_p} \\
    -\sqrt{t_p} & \sqrt{t_s}
\end{array}
\right]
C_j,
\eea
and $t_1 = (t_p -t_s)$, $t_2 = \sqrt{t_s t_p} - t_{sp}$ and $t_3 = \sqrt{t_s t_p} + t_{sp}$.
The wave functions $w_{l/r, \nu} (j)$  of the edge states are introduced by
$b_{l/r} = \sum_{j\, \nu =s,p} w_{l/r,\nu} (j) a_\nu (j)
= \sum_{j\, \alpha =\pm} \tilde{w}_{l/r,\alpha} (j) \phi_\alpha (j)$, 
where $b_{l/r}$ is the fermion operator of the left/right edge state. 
The wave function of the left edge state $b_l$ is given as
\bea
&&\tilde{w}_{l,+} (j) = \left\{
                \begin{array}{cc}
                \exp\left( -j/\xi \right),
                &\text{if $j$ is even};\\
                0, &\text{otherwise,}
                \end{array}
               \right. \nonumber \\
&& \tilde{w}_{l,-} (j) = \left\{
                \begin{array}{cc}
                -\frac{t_1}{t_2+t_3} \exp\left( -j/\xi \right),
                &\text{if $j$ is even};\\
                0, &\text{otherwise,}
                \end{array}
               \right.
\label{eq:constructedwave}
\eea
with the decay width $\xi = \frac{2}{\log(\left| t_3/t_2 \right|)}$.
The wave function of the right edge state can be constructed by performing 
parity transformation of the left edge state.
As long as $t_2$ vanishes, the width $\xi\to 0$ and the edge states
are completely confined on the ends of the ladder. With finite $t_2$ the edge
states deconfine and decays exponentially.
The critical point is $t_2/t_3 =1$ ($t_{sp} =0$), for which $\xi \to \infty$.
At this point the bulk gap closes~(Eq.~(\ref{eq:bulkspectra})).
The analysis presented here is confirmed by numerical calculations.

\section{\bf Mapping to a model of two Majorana fermi chains and deconfined excitations}

In the flat band limit $t_s = t_p = t_{sp} = t$ (or equivalently $t_1 =t_2 = 0$) we can introduce particle-hole mixed 
operators $\phi_{\up/\down} (j) = [a_p (j) \pm a_s ^\dag (j) ] /\sqrt{2}$. The $sp$-orbital 
ladder maps to two decoupled Kitaev's p-wave superconducting chains~\cite{2001_Kitaev_Majorana_Phys-usp}, 
\be
H_0 \rightarrow t 
    \sum_{j \sigma } 
    \left\{ \phi_\sigma ^\dag (j) \phi_\sigma (j+1) 
      +  P_\sigma \phi_\sigma^\dag  (j)  \phi_\sigma ^\dag (j+1) +h.c.\right\},
\ee
with $P_{\up/\down} = \pm$.

It is instructive to rewrite the fermion operators $\phi_\pm$ defined 
in Sec.~\ref{sec:wave-edgestate} in terms 
of Majorana fermion operators as 
$\phi _ \pm = \frac{1}{2} \left( \psi_{1 \pm}  - i \psi_{2\mp}  \right)$, 
with $\psi_{\ell p}   = \psi_{\ell p}  ^\dag$, 
and $\left\{ \psi_{\ell p}  , \psi_{\ell' p'} \right\} = 2 \delta_{p p'} \delta_{\ell \ell'}$.       
The resulting Hamiltonian in Eq.~(\ref{eq:sup-ham}) reads 
\be
H_0  =\frac{i}{2} \sum_j 
    [\psi_{1+}  (j), \psi_{2 -} (j), \psi_{1 -}  (j), \psi_{2 +}  (j)]
\left[
  \begin{array}{cccc}
  0	&-t_1	&0	&-t_2 \\
  t_1	&0	&t_2	&0 \\
  0	&-t_3	&0	&0\\
  t_3	&0	&0	&0
  \end{array}
\right]
\left[
  \begin{array}{c}
  \psi_{1+}  (j+1)\\
  \psi_{2 -}  (j+1)\\
  \psi_{1 -} (j+1)\\
  \psi_{2 +} (j+1)
  \end{array}
\right].	
\ee

In the flat band limit, the Hamiltonian  maps  to  
two decoupled Majorana Fermi chains~\cite{2001_Kitaev_Majorana_Phys-usp}, 
\be
\textstyle H_0 \rightarrow i\frac{t_3}{2} \sum_j \left[
	  \psi_{2 +}  (j) \psi_{1 +}(j+1)
	  -\psi_{1-} (j)  \psi_{2 -} (j+1) \right].
\ee 
The unpaired Majorana fermion operators are 
$\psi_{1+} (0)$, $\psi_{2-}(0)$, $\psi_{2+} (L-1)$ 
and $\psi_{1-}(L-1)$. 
These Majorana fermion operators give a many-body 
ground state manifold of Z$_4$ degeneracy for $\mu =0$ in the grand canonical 
ensemble~\cite{2001_Kitaev_Majorana_Phys-usp}.
Since $[\phi_+ ^\dag (0) \phi_+ (0), H] =0$, 
$[\phi_- ^\dag (L-1) \phi_- (L-1), H] =0$ and 
$[\phi_+ ^\dag (0) \phi_+ (0) ,\phi_-^\dag (L-1) \phi_-(L-1)] =0$, one can 
label the degenerate ground states   
by operators $\phi_+^\dag(0)\phi_+ (0)$  and $\phi_-^\dag (L-1) \phi_- (L-1)$ in such 
a way as follows,
\bea 
&& |G_0 \rangle \text{: } \phi_+ ^\dag (0)  \phi_+ (0) |G_0\rangle=0 
	\text{ and  }
	\phi_- ^\dag (L-1) \phi_- (L-1)  |G_0 \rangle =0, \\
&& |G_+ \rangle = \phi_+ ^\dag (0) |G_0\rangle,\\
&& |G_- \rangle = \phi_- ^\dag (L-1) |G_0 \rangle,\\
&& |G_{+-} \rangle = \phi_+^\dag (0) \phi_- ^\dag (L-1) |G_0\rangle. 
\eea
(The states $|G_+ \rangle$ and $|G_- \rangle$ have the same number of fermions, 
providing the Z$_2$ degeneracy at half filling in the canonical ensemble.)  
In the Z$_4$ degenerate ground state manifold the operators $\phi_+ (0)$ and $\phi_- (L-1)$ read as
\bea
&&\phi_+(0) = |G_0 \rangle \langle G_+ | + |G_-\rangle \langle G_{+-}| \\
&&\phi_-(L-1) = |G_0 \rangle \langle G_- |-|G_+ \rangle \langle G_{+-} |.
\eea
To demonstrate the deconfined Dirac fermion excitations explicitly in this ground state manifold, we 
can label the states by $d_1^\dag d_1$ and $d_2 ^\dag d_2$, where $d_1$ and $d_2$ are fractionalized 
Dirac fermion operators $d_1 = (\psi_{1+} (0) - i \psi_{2 + } (L-1) )/2$ and $d_2  = (\psi_{2-} (0) - i \psi_{1-} (L-1))/2$, respectively. 
The relabeled states are defined by
\bea
&& |{\tilde{G}}_0 \rangle \text{: } d_1 ^\dag d_1 |{\tilde{G}}_0 \rangle=0 
	\text{ and  }
	d_2 ^\dag d_2  |{\tilde{G}}_0 \rangle =0, \\
&& |{\tilde{G}}_1 \rangle = d_1 ^\dag  |{\tilde{G}}_0 \rangle,\\
&& |{\tilde{G}}_2 \rangle = d_2^\dag  |{\tilde{G}}_0 \rangle,\\
&& |{\tilde{G}}_{12} \rangle = d_1^\dag  d_2 ^\dag |{\tilde{G}}_0\rangle. 
\eea 
These states embedding fractionalized Dirac fermions are actually given by 
$|\tilde{G}_0\rangle  = (|G_+\rangle - |{G}_-\rangle)/\sqrt{2}$,
$|\tilde{G}_{1} \rangle = (|G_0 \rangle - |G_{+-} \rangle )/\sqrt{2}$, 
$|\tilde{G}_{2} \rangle = -(|G_0 \rangle + |G_{+-} \rangle )/\sqrt{2}$ 
and $|\tilde{G}_{12} \rangle = -(|G_+  \rangle +|G_- \rangle)/\sqrt{2}$. 
A transition from $|\tilde{G}_0 \rangle$ to $|\tilde{G}_1\rangle$ creates 
one fractionalized Dirac fermion excitation.

Away from the flat band limit the unpaired Majorana fermion operators 
$\psi$  are defined by $[\psi, H_0] =0$. The four unpaired Majorana fermion operators 
are $\psi_1 = b_l   + b_l ^\dag $, $\psi_2 = i(b_l -b_l ^\dag )$, 
$\psi_3 = b_r + b_r ^\dag$  and $\psi_4 = i(b_r - b_r ^\dag )$, where the Dirac fermion operators 
$b_l$ and $b_r$ are defined in Sec.~\ref{sec:wave-edgestate}.

\section{\bf The gapped interpolation.}
An interpolation between $\mathcal{H} (k)$ and $\mathcal{H}'(k)$ is defined as
$h(k, \varphi) = \mathcal{H} (k) +
    \frac{\Delta_y}{2} \left[1-\cos(\varphi) \right]\sigma_y +
    \Delta_x \sin (\varphi)  \sigma_x
= d_0 (k) \mathbb{bb} + \vec{d}(k, \varphi) \cdot \vec{\sigma}$,
with $h(k, 0) = \mathcal{H} (k) $ and $h(k, \pi) = \mathcal{H}' (k)$.
The eigenvalues of this interpolation exhibit a finite
gap for any ($k,\varphi$), given a sufficiently large $\Delta_x$.
The Chern number $C_1$ defined by
$ C_1  = \frac{1}{4\pi} \int dk d\varphi \hat{d} \cdot
    \frac{\partial \hat{d} }{\partial k}
    \times \frac{\partial \hat{d} }{\partial \varphi}
$
is $1$ for $\Delta_y > \Delta_y ^c$, and is $0$ for $0<\Delta_y<\Delta_y ^c$.
Since the Chern parity $(-1)^{C_1}$ is well defined for gapped
interpolations with fixing $h(k, 0)$ and $h(k, \pi)$, $C_1$
being odd tells $\mathcal{H} (k) $ is topologically distinguishable
from $\mathcal{H}' (k)$~\cite{2008_Qi_TFT_TI}.

\section{\bf Effective field theory of the Domain Wall}
To calculate the charge modulation induced by a domain wall connecting a topological insulator and a
trivial insulator, we couple the ladder system to an auxiliary charge $U(1)$ gauge field $(A_t, A_x)$
and construct a low energy effective field theory, which is $1+1$ dimensional reduced Chern-Simons Field theory~\cite{2008_Qi_TFT_TI}.
A ${\eta}$ dependent model Hamiltonian describing the domain wall is introduced
\bea
H_{\eta} &=& H_0 + \frac{\Delta_y}{2}\sum_j
    \left[
        1- \cos({\eta} )
    \right] C_j ^\dag \sigma_y C_j
     + \Delta_x \sin({\eta}) C_j ^\dag \sigma_x C_j \\
     &=& \sum_k \tilde{C}_k \mathcal{H}_{\eta} (k) \tilde{C}_k.
\eea
Here the $\sigma_x$ term is introduced to guarantee a finite energy gap ($\Delta_\epsilon$) of $H_{\eta}$.
We split the ${\eta}$ field into two parts as ${\eta} = {\eta}_0 + \delta{\eta}(j,\tau)$.
The field ${\eta}_0$ satisfies the boundary condition---
${\eta}_0(-\infty,\tau) = 0$, ${\eta}_0 (\infty,\tau) = \pi$. For
example it can be set as
${\eta}_0(j,\tau)= \arctan \left(\frac{j}{l_{dw} }\right) + \frac{\pi}{2}$, where
$l_{dw}$ is the characteristic width of the domain wall. To proceed we assume $l_{dw}$ is much
larger than the microscopic length scale ($\sim \frac{1}{\Delta_\epsilon ^{1/z } }$) of the fermion system.
The field $\delta {\eta}(j,\tau)$ satisfies a periodic boundary condition.
In the following derivation we will treat ${\eta}_0$ as a quasi-static field since
it varies slowly in space. (One can think that we are deriving a local effective
field theory for a subsystem in which the variation of ${\eta}_0$ is negligible.)

With the model Hamiltonian $H_{\eta}$ coupled to the gauge field,
its Lagrangian is given by 
\bea
&&L[C^\dag, C] =
\int_0 ^\beta d\tau \mathcal{L} (\tau) \nonumber \\
&& \mathcal{L} = \sum_j
    \left\{
        C_j ^\dag D_\tau C_j
        + C_j ^\dag \mathcal{T}_{j\, j+1} C_{j+1}
        +C_{j+1} ^\dag \mathcal{T}_{j+1\, j} C_j
        + \frac{\Delta_y}{2} \left[1-\cos({\eta}) \right]
        C_j ^\dag \sigma_y C_j
        + \Delta_x \sin({\eta}) C_j ^\dag \sigma_x C_j
    \right\},
\eea
with
\bea
D_\tau &=& \partial_\tau + i A_\tau (j, \tau), \nonumber \\
\mathcal{T}_{j\,j+1} &=& e^{i A_x (j+1/2, \tau) } T_{j\,j+1} ,\nonumber \\
T_{j\,j+1} &=&  \left[
                \begin{array}{cc}
                -t_s & -t_{sp} \\
                t_{sp} & t_p
                \end{array}
                \right], \nonumber \\
\mathcal{T}_{j\,j+1} &=& \mathcal{T}_{j+1\,j} ^\dag.
\eea
It is readily verified that the Lagrangian has a $U(1)$ gauge symmetry defined by
\bea
&& A_\tau \to A_\tau +\partial_\tau \vartheta (j,\tau) \nonumber \\
&& A_x (j+1/2, \tau) \to A_x (j+1/2, \tau) +
            \left[
                \vartheta (j+1,\tau) - \vartheta(j, \tau)
            \right] \nonumber \\
&& C(j,\tau) \to C(j,\tau) e^{-i\vartheta(j,\tau)}.
\eea
In the continuum limit, $A_x$ transforms as
$A_x (x,\tau) \to A_x (x,\tau) + \partial_x \varphi (x,\tau)$.
In the following we shall write
$k$ as $k_x$, ${\eta}_0$ as $k_y$ and $\delta {\eta}$ as $A_y$
just to make the equations more compact.

In the momentum-frequency ($K \equiv (k_x,i\omega)$) space the Lagrangian reads
\bea
L &=& \sum_{K K'} \tilde{C}^\dag (K) \mathcal{D}_{K K'} \tilde{C}(K'), \nonumber \\
\mathcal{D} &=& \mathcal{D}^{(0)} + \delta \mathcal{D},\nonumber \\
\mathcal{D} ^{(0)} _{K K'} &=& (-i\omega + \mathcal{H}_{\eta} (k_{x}) ) \delta_{K K'}  \nonumber \\
\delta \mathcal{D} _{K K'} &=& \sum_\nu A_\nu (K -K') \Gamma_\nu (\frac{k_x +k_x '}{2}),
\eea
where $\Gamma_\tau = i\mathbb{I}$,
$\Gamma_{\alpha=x,y} =\frac{\partial \mathcal{H}_{\eta} (k) }{\partial k_\alpha}$.
For convenience, we further introduce $G(K) = \left[-i \omega + \mathcal{H}_{\eta} (k)\right]^{-1}$

The effective action of the gauge field $A_\nu$ is defined as
\bea
S_{\text{eff}} [A_\nu] = -\ln \int D[C^\dag, C] \exp\left(- L[C^\dag, C; A_\nu]\right),
\eea
which can be calculated order by order as follows.
\bea
S_{\text{eff}}[A_\nu] &=& -\ln \det [\mathcal{D}] = -\text{Tr} \ln \mathcal{D}
=- \text{Tr} \ln \left[\mathcal{D}^{(0)} + \delta \mathcal{D} \right]\nonumber \\
&=& -\text{Tr} \ln \mathcal{D} ^{(0)}
    -\text{Tr} \left[ \mathcal{D}^{(0)\, -1}\delta \mathcal{D} \right]
    +\frac{1}{2} \text{Tr} \left[\mathcal{D}^{(0)\, -1}\delta \mathcal{D} \right]^2 +\ldots.
\eea
Since we are calculating the charge/current induced by the domain wall configuration---${\eta}_0$ field, we only 
keep the quadratic terms and thus the effective action is given by 
\bea
S_\text{eff} &=& \frac{1}{2} \text{Tr} \left[\mathcal{D}^{(0)\, -1}\delta \mathcal{D} \right]^2 \nonumber \\
&=& \frac{1}{2}  \sum_{\mu\nu Q} A_\mu (Q) A_\nu (-Q) \text{Tr} \sum_{K} G(K+Q/2) \Gamma_\mu (K) G(K-Q/2) \Gamma_\nu (K).
\eea
Although the theory is complicated, it can be simplified because we are only interested in a low energy theory.
The generic form of $S_\text{eff}$ at low energy is
\bea
S_{\text{eff}} = \beta L \sum_K A_\mu (Q) A_\nu (-Q) \left( q \mathcal{K}^{\mu \nu} + i q_0 \mathcal{W}^{\mu \nu}\right),
\eea
with $Q = (q, iq_0)$.
The massive terms are prohibited due to gauge invariance. 
By comparing the above equations we get
\bea
\mathcal{W}^{\mu \nu} &=& \frac{1}{2 \beta L} \lim_{Q \to 0}\frac{d}{d (iq_0)}
    \text{Tr} \sum_{K} G(K+Q/2) \Gamma_\mu (K) G(K-Q/2) \Gamma_\nu (K), \\
\mathcal{K}^{\mu \nu} &=& \frac{1}{2\beta L}\lim_{Q \to 0} \frac{d}{d q}
    \text{Tr} \sum_{K} G(K+Q/2) \Gamma_\mu (K) G(K-Q/2) \Gamma_\nu (K).
\eea
Then we have
\bea
\mathcal{W} ^{\mu \nu} &=& \frac{1}{4 \beta L}
        \text{Tr} \sum_K \left[ \partial_{i\omega} G(K)\right] \Gamma_\mu (K) G(K) \Gamma_\nu (K) - \mu\leftrightarrow \nu , \\
\mathcal{K} ^{\mu \nu} &=& \frac{1}{4 \beta L}
        \text{Tr} \sum_K \left[ \partial_{k} G(K)\right] \Gamma_\mu (K) G(K) \Gamma_\nu (K) - \mu\leftrightarrow \nu.
\eea
It is clear that $\mathcal{W}$ and $\mathcal{K}$ are anti-symmetric.
From Ref.~\cite{2008_Qi_TFT_TI},
$
\mathcal{W} ^{xy} = \frac{i}{2} \oint \frac{d k_x}{2\pi} \Omega_{ k_x{\eta} } = \frac{i}{4\pi} \partial_{\eta} \gamma ({\eta})
$
where $\Omega_{k_x {\eta}}$ ($\gamma({\eta})$) is the Berry curvature (Berry phase) of the Hamiltonian $\mathcal{H}_{\eta}$.
Other terms of $\mathcal{W}$---$\mathcal{W}^{\tau \nu}$---vanish because $G(K)$ and $\partial_{i\omega} G(K)$ commute.
The commuting relation can be verified by choosing the eigen-basis of $\mathcal{H}_{\eta}(k)$.
Now we have fully established the frequency part of the effective action, which is
\bea
\beta L \sum_K A_\mu (K) A_\nu (K) i\omega \mathcal{W} ^{\mu \nu}
=2\beta L\sum_K W^{xy} A_x (-K) (-i\omega) A_y (K)
\eea
In the real space (continuum limit) this term reads 
$
2 \int d x d \tau \mathcal{W} ^{yx} (A_y \partial_\tau A_x).
$
From gauge invariance, it is readily proved that $-i \mathcal{K} ^{\tau y} = \mathcal{W}^{xy}$.
Thus the effective action is given by
\bea
S_\text{eff} [A_\mu] =2 \mathcal{W}^{yx} \int d x d \tau (A_y \partial_\tau A_x - A_y \partial_x A_\tau).
\eea
The  real time action reads
\bea 
\tilde{S}_\text{eff} [A_\mu] &=&
      -2 i \mathcal{W}^{yx}\int dx d t (A_y \partial_t A_x - A_y \partial_x A_t) \nonumber \\
    &=& \int dx dt (A_x \partial_t {\eta} - A_t \partial_x {\eta} ) \frac{1}{2\pi} \partial_{\eta} \gamma({\eta}) .
\eea
We have replaced $\partial_\nu \delta {\eta}$ by $\partial_\nu {\eta}$ because $\partial_\nu {\eta} \approx \partial_\nu \delta {\eta}$. 
This action can actually be written down directly from the famous $2D$ Chern-Simons field theory by a dimension reduction procedure~\cite{2008_Qi_TFT_TI}.
The linear response of charge/current is given by 
$j_\nu = \frac{\delta S_{\text{eff}}} {\delta A_\nu}$.
The charge carried by the domain wall is given by 
$Q = \int \frac{\tilde{S}_\text{eff}} {\delta A_t} = -\frac{1}{2\pi} \int dx \partial_x {\eta} \partial_{\eta} \gamma({\eta}) 
= - \int \frac{d{\eta}}{2\pi} \partial_{\eta} \gamma({\eta})$. The charge $Q$ is $1/2$ with $\Delta_x >0$. The microscopic details of the 
domain wall can change $Q$ by $1$, so we conclude the charge carried by the domain wall is $Q = \frac{1}{2} \mod 1$.  

\section{\bf Local density fluctuation of topological (non)-trivial insulators}
In this section we derive an analytic formula for the density fluctuation, which can be measured in cold atom experiments. 
The density fluctuation in cold atom experiments measures 
$\delta \rho \equiv \frac{1}{L} \sum_j\left \langle\left( C_j ^\dag C_j -1\right)^2 \right \rangle $.

First two flat-band limits---(1) ($\Delta_y =0$, $t_s = t_p = t_{sp}\neq 0$) and
(2) ($\Delta_y \neq 0$, $t_s = t_p = t_{sp} =0$) are explored. The two limits give different Berry phases ($\gamma = \pi$
for Limit (1) and $\gamma = 0$ for Limit (2)). Limits (1) and (2) give non-trivial and trivial insulators respectively.  
For limit (1) fermions live on the bonds (one fermion per bond). The ground state can be written as
\bea
|G\rangle = \prod_j \frac{1}{\sqrt{2}} \left( \phi_- ^\dag  (j) -\phi_+ ^\dag (j+1) \right) |0\rangle,
\eea
with $|0 \rangle $ the vacuum state with no particles. 
$\langle (C_j ^\dag  C_j)^2 \rangle$ is given as follows
\bea
\langle (C_j ^\dag  C_j)^2 \rangle 
 &=& \sum_ {\alpha = \pm, \alpha' = \pm}
    \langle \phi_\alpha ^\dag (j) \phi_\alpha (j) \phi_{\alpha'} ^\dag (j) \phi_{\alpha'} (j) \rangle \nonumber \\
&=&  \sum_{\alpha, \alpha'} \left\{
                \delta_{\alpha \alpha'} \langle \phi_\alpha ^\dag  (j)  \phi_{\alpha'} (j) \rangle
                -\delta_{\alpha, -\alpha'}
                    \langle \phi_\alpha ^\dag  (j) \phi_{\alpha'} ^\dag  (j)\phi_{\alpha} (j)  \phi_{\alpha'} (j) \rangle
                    \right\} \nonumber \\
&= & \sum_\alpha \left\{
                    \langle \phi_\alpha ^\dag (j)  \phi_\alpha (j) \rangle
                    - \langle \phi_\alpha ^\dag  (j) \phi_{-\alpha} ^\dag  (j)\phi_{\alpha} (j)  \phi_{-\alpha} (j) \rangle
                \right\}               \nonumber \\
&= & 1 + 2\times \left(\frac{1}{\sqrt{2}}\right)^4 = \frac{3}{2}
\eea
Then we have $\delta \rho = \frac{1}{\sqrt{2}}$.
For Limit (2) there is no hopping between different unit cells and fermions localize on each unit cell. 
Fermions cannot tunnel between different unit cells, so the local density fluctuation vanishes, i.e., $\delta \rho =0$.

For the generic case, the periodic boundary condition is adopted 
{ and the calculation is performed in the momentum space. In momentum space, 
the Fourier transformed operators are defined as 
$\tilde{\alpha}( k) = \frac{1}{\sqrt{L}} \sum_j \alpha_j e^{-i kj}$.} 
The density fluctuation, which is 
a bulk property, does not depend on the boundary condition in the thermodynamic limit. 
The fermion operators of the eigen-modes (labeled by $k$) are introduced by
$
\left[
    \tilde{b}_\up (k) ,
    \tilde{b}_\down (k)
  \right]^T =
    U^\dag \tilde{C}_k 
$. 
$\up/\down$ here means the upper/lower band. 
The unitary matrix $U$ is defined in Sec.~\ref{sec:top-index}. 
The ground state of the fermionic ladder at half filling is
$
|G \rangle = \prod_k \tilde{b}_\down ^\dag (k) |0\rangle.
$
The calculation of $\langle (C_j ^\dag C_j )^2 \rangle$ is as follows
\bea
&& \langle (C_j ^\dag C_j )^2 \rangle\nonumber \\
= && \frac{1}{L^2} \sum_{\nu=s/p, \nu'=s/p} \sum_{k_1 k_2 k_3 k_4} e^{ i(k_2 + k_4 - k_3 -k_1)j}
    \langle \tilde{a}_\nu ^\dag (k_1) \tilde{a}_ \nu (k_2) \tilde{a}_{\nu'} ^\dag (k_3) \tilde{a}_{\nu'} (k_4) \rangle
    \nonumber \\
= && \frac{1}{L^2} \sum_{\nu, \nu'} \sum_{k k'}
    \left\{
        \langle   \tilde{a}_\nu ^\dag (k) \tilde{a}_ \nu (k) \rangle \langle \tilde{a}_{\nu'} ^\dag (k' ) \tilde{a}_{\nu'} (k') \rangle
        + \langle  \tilde{a}_\nu ^\dag (k)\tilde{a}_{\nu'} (k) \rangle \langle \tilde{a}_ \nu (k') \tilde{a}_{\nu'} ^\dag (k')  \rangle
    \right\}\nonumber \\
= && \frac{1}{L^2} \sum_{k k'}
    \left\{
        \langle\tilde{b}_\down ^\dag (k)\tilde{b}_\down (k)\tilde{b}_\down ^\dag (k')\tilde{b}_\down (k') \rangle
        + \sum_{\nu \nu',s_1 s_2 s_3 s_4} \langle \tilde{b}_{s_1} ^\dag (k) \tilde{b}_{s_2} (k') \tilde{b}_{s_3} ^\dag (k') \tilde{b}_{s_4} (k) \rangle
                [U (k) ]^* _{\nu s_1} [U (k') ]_{\nu s_2} [U (k') ] ^* _{\nu' s_3} [U (k)] _{\nu' s_4}
    \right\} \nonumber \\
= &&\frac{1}{L^2} \sum_{k k'} \left\{ 1
            +\langle \tilde{b}_\down ^\dag (k) \tilde{b}_\up(k') \tilde{b}_\up ^\dag (k') \tilde{b}_\down (k) \rangle
            [U^\dag (k) ] _{ \down \nu } [U (k') ]_{\nu \up } [U ^\dag (k') ] _{\up \nu'} [U (k)] _{\nu' \down}
            \right\} \nonumber \\
= && 1 + \frac{1}{L^2} \sum_{k k'} | [U^\dag (k) U (k')  ]_{\down \up} |^2  .
\eea
The term $\frac{1}{L^2} \sum_{k k'} | [U^\dag (k) U (k')  ]_{\down \up} |^2$ simplifies as
\bea
&&\frac{1}{L^2} \sum_{k k'} \left| [U^\dag (k) U (k')  ]_{\down \up} \right|^2 
=\frac{1}{L^2} \sum_{k k'} \left| \left[e^{-i\sigma_x \theta (k)/2} e^{i \sigma_x \theta (k')/2} \right]_{\down \up} \right|^2 
= \frac{1}{L^2} \sum_{k k'} \sin^2 \left( \frac{\theta (k) - \theta(k')}{2} \right) \nonumber \\
&=& \frac{1}{2} \left[ 1 - \left(\oint \frac{dk}{2\pi} \cos (\theta (k)) \right)^2
                - \left(\oint \frac{dk}{2\pi} \sin(\theta(k)) \right)^2 \right].
\eea
Due to the particle-hole symmetry $h_z(k) = -h_z (\pi -k)$ and that $h(k) = h(\pi-k)$,
and  we thus have $\cos (\theta(k)) = -\cos(\theta(\pi - k))$, so that 
 $\oint dk \cos(\theta(k)) =0$. 
Then the density fluctuation of a particle-hole symmetric insulator is given by 
\bea
\delta \rho^2 = \frac{1}{2} \left[ 1 - \left(\oint \frac{dk}{2\pi} \sin(\theta(k)) \right)^2 \right].
\eea

In the absence of the the imaginary transverse tunneling ($\Delta_y =0$), time-reversal symmetry is also respected. 
Here we have  $\sin(\theta(k)) = - \sin(\theta(-k))$ because  $h_y (k) = -h_y(-k)$.
Apparently $\oint dk \sin(\theta(k))=0$, so we conclude 
$\delta \rho^2  = \frac{1}{2}$  for $\Delta_y =0$, regardless of $t_s$, $t_p$ and $t_{sp}$.


\section{\bf The Ferro-Orbital order of the Mott insulator}
The quantum phase in the strong interaction regime $U_{sp}>U_{sp} ^c$ is a Mott insulator.
The double occupancy in this phase is found to be greatly suppressed, i.e., $\langle n_s (j) n_p (j)\rangle \ll 1$.
The orbital physics of this Mott insulator can be described
by an effective Hamiltonian with the double occupancy projected out. Then the two states,
$a_s^\dag (j) |\Omega \rangle$ and $(-1)^j a_p ^\dag (j) |\Omega \rangle$, are mapped 
to two pseudo-spin $\frac{1}{2}$ states. The resulting effective Hamiltonian 
(for $t_s = t_p \equiv t_0$) is the well-known
$XXZ$ Hamiltonian given as
\bea
 H_\text{eff} &=&  \sum_{<ij>}
            \left\{J_{yz} \left(\mathbf{S}_y (i) \mathbf{S}_y(j) + \mathbf{S}_z(i) \mathbf{S}_z(j) \right)
               + J_{x} \mathbf{S}_x (i) \mathbf{S}_x(j)\right\},
\eea
with
$J_{x} = 2\frac{ t_0 ^2 + t_{sp} ^2} {U_{sp}}$ and $J_{yz} = 2 \frac{t_0 ^2 - t_{sp} ^2}{U_{sp}}$.
The $XXZ$ Hamiltonian predicts a gapped Mott insulator for the 
$sp$-orbital ladder with a ferro-orbital order $\langle C_j ^\dag \sigma_x C_j\rangle$, 
which spontaneously breaks the particle-hole symmetry.

\section{\bf The stability against inter-ladder coupling}
In this section the stability of the topological property of the $sp$-orbital ladder against the 
inter-ladder coupling will be proved. 
Due to the experimental setup the leading inter-ladder coupling is the coupling between the $B$ ($A$) chain 
and the $A$ ($B$) chain of the nearest ladder. The tight binding Hamiltonian describing such a coupled 
two dimensional system is given as
\bea
H_{2D} = \sum_{\tbf{R}} 
      \left\{ C_\tbf{R} ^\dag T C_{\tbf{R} + \hat{x}} 
      + C_{\tbf{R}} ^\dag T' C_{\tbf{R} + \hat{x} + \hat{y}} 
      -C_{\tbf{R} } ^\dag T' C_{\tbf{R} + \hat{y} -\hat{x}} 
  +h.c. \right\},
\eea
where $\tbf{R}$ labels the positions of lattice sites and $\hat{x}$ ($\hat{y}$) 
is the primitive vector in the $x$- ($y$-) direction. The inter-ladder coupling matrix $T'$ is given 
as 
$
\left[
  \begin{array}{cc}
   0 & 0 \\
   t_{sp} ' &0
  \end{array}
\right],
$  
with $t_{sp} '$ the inter-ladder coupling strength. 
In the momentum space the Hamiltonian reads 
$
H_{2D}  = \sum_\tbf{k} \tilde{C} ^\dag (\tbf{k}) \mathcal{H}_{2D} (\tbf{k}) \tilde{C} (\tbf{k})  
$, with 
\bea
\mathcal{H} _{2D} (\tbf{k}) = 
  \left[
  \begin{array}{cc}
   -2 t_s \cos k_x &	-2i(t_{sp} + t_{sp} ' e^{-ik_y} ) \sin k_x \\
  2i(t_{sp} + t_{sp} ' e^{ik_y} ) \sin k_x & 2 t_p \cos k_x 
  \end{array}
  \right]. 
\eea
The Hamiltonian $\mathcal{H}_{2D} (\tbf{k}) $ can be rewritten as 
$\mathcal{H}_{2D} (\tbf{k}) = U^\dag (k_y) \tilde{\mathcal{H}}_{2D} (k_x, k_y)  U(k_y)$ 
with  
\bea
\tilde{\mathcal{H}}_{2D}  (k_x, k_y)=  
  \left[
  \begin{array}{cc}
  -2 t_s \cos k_x &	-2i\tilde{t}_{sp} \sin k_x \\
  2i \tilde{t}_{sp} \sin k_x & 2 t_p \cos k_x 
  \end{array}
  \right] 
\eea
and $ U(k_y) = e^{i\sigma_z \varsigma (k_y)/2}$, 
where 
$\tilde{t}_{sp} = | t_{sp} + t_{sp} ' e^{-ik_y}|$, 
and $\varsigma (k_y) = \arg (t_{sp} +t _{sp} ' e^{-ik_y})$.  
With $k_y$ fixed, $\mathcal{H}_{2D} (k_x, k_y)$ defines a one dimensional system. 
Because the form of  $\tilde{\mathcal{H}}_{2D} (k_x, k_y)$ is exactly the same as 
the one dimensional Hamiltonian describing the $sp$-orbital ladder in the limit 
of $t_{sp} ' =0$, the Berry phase  of $\mathcal{H}_{2D} (k_x, k_y)$ for any given $k_y$ 
is $\pi$. We thus conclude that the one dimensional topological insulator survives even 
with finite inter-ladder couplings. 

The existence of edge states in presence of inter-ladder coupling is verified by 
directly calculating the energy spectra of the 2D system on a cylinder geometry (an 
open cylinder in the $x$-direction). From the energy spectra shown in 
Fig.~\ref{fig:2d_flat_edge} it is clear that the edge states are stable 
against inter-ladder couplings.

\end{widetext}

\bibliographystyle{naturemag}
\bibliography{1Dsp}

\end{document}